# Shubnikov–de Haas Oscillations and Nontrivial Topological State in a New Weyl Semimetal Candidate SmAlSi

Longmeng Xu[1,2,*], Haoyu Niu[1,2,*], Yuming Bai[1,2], Haipeng Zhu[1,2], Songliu Yuan[2], Xiong He[1], Yang Yang[3], Zhengcai Xia[1,2,a)], Lingxiao Zhao[1,2,a)], Zhaoming Tian[1,2,a)]

[1]Wuhan National High Magnetic Field Center, Huazhong University of Science and Technology, Wuhan, 430074, P. R. China

[2]School of Physics, Huazhong University of Science and Technology, Wuhan, 430074, P. R. China

[3]School of Physics and Electronic Engineering, Zhengzhou University of Light Industry, Zhengzhou 450002, P. R. China

[*]These authors contributed equally to this work.

E-mail: xia9020@hust.edu.cn, zhaolx@hust.edu.cn, tianzhaoming@hust.edu.cn

**ABSTRACT:** We perform the quantum magnetotransport measurements and first-principles calculations on high quality single crystals of SmAlSi, a new topological Weyl semimetal candidate. At low temperatures, SmAlSi exhibits large non-saturated magnetoresistance (MR) ~5200% (at 2 K, 48 T) and prominent Shubnikov–de Haas (SdH) oscillations, where MRs follow the power-law field dependence MR$\propto\mu_0H^m$ with exponent m~1.52 at low fields ($\mu_0H$ < 15 T) and linear behavior m~1 under high fields ($\mu_0H$ > 18 T). The analysis of angle dependent SdH oscillations reveal two fundamental frequencies originated from the Fermi surface (FS) pockets with non-trivial π Berry phases, small cyclotron mass and electron-hole compensation with high mobility ( $\mu_\mathrm{h} = 9735 \cdot V^{-1} \cdot s^{-1}$ and $\mu_\mathrm{e} = 2195 \mathrm{cm}^2 \cdot V^{-1} \cdot s^{-1}$ ) at 2 K. In combination with the calculated nontrivial electronic band structure, SmAlSi is proposed to be a paradigm for understanding the Weyl fermions in the topological materials.

Weyl semimetal (WSM) as a new type of quantum state of matter hosting low energy relativistic quasiparticles, has attracted significant attention for both scientific community and potential quantum device applications.[1-4] The typical feature related to the exotic quasiparticles is the topologically protected linear crossings called Weyl nodes near Fermi surface,[5,6] which come in pairs with a definite chirality. Generally, there are two approaches to generate Weyl semimetal, either by breaking space-inversion (SI) or time-reversal (TR) symmetry. The former WSM states have been well studied in various types of nonmagnetic materials with specific crystalline symmetry since its initial discovery in TaAs structural family.[7,8] The latter named as magnetic WSMs have been verified experimentally in a handful of materials, such as $Mn_3X(X=Sn,Ge)$,[9,10] $Co_3Sn_2S_2$,[11,12] $Fe_3Sn_2$[13] and $Co_2MnGa$.[14] While, magnetic WSMs can exhibit unique quantum transport phenomena despite the extremely large linear magnetoresistance (MR)[15] and chiral-anomaly-induced negative MR in nonmagnetic WSM,[16] indeed the large anomalous Hall conductivity (AHC) and topological Hall effect can appear even at zero field originating from the interplay between diverse magnetism and nontrivial Weyl band topology.[12,17] Experimentally, quantum transport study is an important approach to identify the WSM state, which can extract information associated with the topological characteristics of WSM, such as nontrivial π Berry phase in the Shubnikov–de Haas (SdH) oscillations,[7,8] large linear MR,[15,18] small cyclotron mass[7,19] and high carrier mobility.[3,5] In other words, these signatures provide the direction for exploring new topological WSM.

Recently, rare-earth based RAlX compounds (R is light rare-earth element; X is Si or Ge) with noncentrosymmetric space group I4$_1$md have been predicted to host various WSM states including type-I and/or type-2 Weyl fermions by choice of R ions,[20,21] and the topological characteristics of Weyl fermions have been detected by angle-resolved photoemission spectroscopy (ARPES) and first-principles calculations. More importantly, this family provides a rare system for comparative study on nonmagnetic WSM (R=La) and magnetic WSM (R=Pr-Sm) while keeping the same crystal structure, it also enables the tunability of conduction electron of Weyl nodes with different magnetic ground states and anisotropic magnetic behaviors by varying rare-earth ions, like the easy *c*-axis in ferromagnetic (FM) PrAlX[22,23] and easy-plane anisotropy of antiferromagnetic (AFM)-ordered CeAlX,[24,25] etc. Even though the SdH oscillations were studied in RAlSi (R=Pr, Nd),[23,26] experimental identification for existence of Weyl fermions are still insufficient and under debate, then electronic topology in other RAlX members is needed to disclose the underlying Weyl physics. Among this family, SmAlSi has smaller unit-cell parameters compared to previously studied RAlX(R=La-Nd) compounds, and it is AFM order with Neel temperature ($T_N$) ~11 K and effective magnetic moments $\mu_{eff}$~0.75 $\mu_B$/Sm (see the Figure S1 in Supporting Information for details). Therefore, SmAlSi is another suitable system for studying topological properties in the RAlX family.

In this work, we present a systematic magnetotransport studies on high quality single crystals of SmAlSi to explore its nontrivial topological state. Low temperature magnetotransport reveal the existence of large non-saturated MRs ~5200% at 2 K up to ~48 T and quantum SdH oscillations. Moreover, the analysis of SdH oscillations unveil the nonzero Berry phase, light cyclotron mass and high carrier mobility as experimental signatures of Weyl fermions in SmAlSi, in consistent with our first-principles calculations.

SmAlSi single crystals were grown by a self-flux method,[22] high-purity Sm (99.9%, Alfa Aesar), Si (99.99%, Alfa Aesar) and Al (99.9%, Alfa Aesar) pieces were used as starting materials. The ingredient ratio with Sm:Si:Al=1:1:10 were weighted and sealed inside a crucible under vacuum in quartz tube. The loaded quartz tube was cooled slowly from 1150 °C to 800 °C, then was taken out from the furnace and decanted by a centrifuge to remove excess Al flux, large plates of SmAlSi single crystals were obtained with typical dimension 3mm × 2mm × 1mm (see the inset of Figure 1c). To identify the crystal structure of SmAlSi, powder and single-crystal x-ray diffraction (XRD) at room temperature were characterized by using a Rigaku x-ray diffractometer with Cu Kα radiation and analyzed by Rietveld method. Low field electrical transport measurements were carried out with a standard four-probe technique in commercial Physical Property Measurement System (PPMS, 14 T) using samples with a rectangle shape. To eliminate the influence of misalignment of electrodes on Hall resistivity, it was measured at both positive and negative fields and antisymmetrized by $\rho_{yx}(H)= [\rho_{yx}(+H)-\rho_{yx}(-H)]/2$. High-field magnetotransport were measured in Wuhan National High magnetic field Center with pulsed magnetic field up to 48 T. The electronic structures were calculated by including the spin-orbital coupling (SOC) using the projector-augmented wave (PAW) method[27] as implemented in Vienna Ab-initio Simulation Package (VASP).[28] The exchange-correlation were included using the Perdew-Burke-Ernzerhof (PBE).[29,30] After convergence test, a 500 eV energy cut-off was used, and self-consistent cycles were proceeded on a 12×12×12 Monkhorst-Pack k-point mesh.

The powder XRD profiles can be well refined by the Rietveld method with reliability parameters $R_p$ = 3.78%, $R_{wp}$ = 5.46%, and $\chi^2$ = 1.26, manifesting high-quality SmAlSi crystallized into the tetragonal structure with a noncentrosymmetric space group $I4_1md$ (No.109) (see Figure 1b). The structural lattice parameters and atomic coordinates are summarized in Table S1 (see Supporting Information for details). Notably, SmAlSi has smallest unit-cell lattice parameters $a = b$ = 4.1586 Å and $c$ = 14.4332 Å among the serial RAlSi(Ge) family members with R = La-Sm.[20-26] Figure 1c shows the single crystal XRD pattern of SmAlSi, indicating that the surface of crystal is $ab$ plane and $c$-axis perpendicular to the plates.

Figure 2a shows the temperature ($T$) dependence of longitudinal resistivity $\rho_{xx}(T)$ under different fields. During the measurement, electrical current ($I$) is along $a$-axis ($I // a$) and field ($\mu_0 H$) is parallel to $c$-axis ($\mu_0 H // c$). Under zero field, SmAlSi exhibits a typical metallic behavior with residual resistivity ratio RRR = $\rho_{xx}(300K)/\rho_{xx}(2K)$ = 5.5, this value is largest among the reported values including other sister compounds in RAIX family.[20-26] As increased fields, $\rho_{xx}(T)$ is enhanced gradually together with the occurrence of broad humps ($T_1$) and valleys ($T_2$) at $T > T_N$, where they simultaneously appear as $\mu_0 H \geq 6$ T. For $\mu_0 H \leq 2$ T, $\rho_{xx}(T)$ exhibits metallic behavior with two cusp-like anomalies connected to the magnetic transitions (see Figure S1a in the Supporting Information), followed by a downward trend. At high fields ($\mu_0 H \geq 4$ T), $\rho_{xx}(T)$ curves change to an upturn profile below $T_N$. Based on the $\rho_{xx}(T)$ and magnetic results, we constructed the temperature-field phase diagrams shown in Figure 2b. To highlight the evolution of dominant conduction mechanisms, the contour plot of $d\rho/dT$ is presented. At low temperatures ($T<T_N$), $d\rho/dT > 0$ gradually changes to $d\rho/dT < 0$ as increased field, indicative of the existence of correlation between magnetic structure and electron conductivity. In paramagnetic state (PM) above $T_N$, a crossover from metallic ($d\rho/dT > 0$) to semiconducting ($d\rho/dT < 0$) behavior also happens, which is separated by $T_1$ or $T_2$. This transition can be attributed to the multiband effect as observed in other semimetals with electron-hole compensation[31] or to an excitonic gap induced by magnetic field.[32] Overall, the above field-temperature phase diagram implies the electronic state near Fermi level is sensitive to external magnetic field.

The high-field MRs defined as MR $= [\rho_{xx}(\mu_0 H) - \rho_{xx}(0)]/\rho_{xx}(0) \times 100\%$ are shown in Figure 2c. At low temperatures, SmAlSi exhibits an extremely large non-saturated MR behavior, as an example, MR reaches ~ 5200% under $\mu_0 H$ = 48 T at 2 K. To clarify this field dependent MR behavior, a double-logarithmic plot of MR versus $\mu_0 H$ is shown in Figure 2d. As seen, MR follows power-law field dependence MR$\propto \mu_0 H^m$ with exponent m~1.52 at low fields ($\mu_0 H$ < 15 T) and cross over to linear behavior with m~1 for high fields ($\mu_0 H$ > 18 T). This field dependence is different from the quadratic field dependence predicted by the two-band theory for semimetal with balanced electrons and holes as reported in WTe$_2$[33] and rare-earth monopnictides.[34,35] On other side, high-field linear dependent MR may be derived from the linear dispersive structures as report in the Dirac/Weyl materials such as in Cd$_3$As$_2$[15] and TaP.[36] While, the electron-hole resonance mechanism can't be completely excluded since two type carries are coexistent as later discussed on Hall effect.

Figure 3a presents the Hall resistivity $\rho_{yx}(\mu_0 H)$ at different temperatures under the *I* // *a* and $\mu_0 H$ // *c* configurations for measurement. At 2 K, $\rho_{yx}(\mu_0 H)$ shows a nonlinear behavior, both $\rho_{yx}(\mu_0 H)$ and its slope change sign from positive at low fields to negative at high fields, supporting the two-type carriers coexistent in SmAlSi. As increased temperatures, the slopes of $\rho_{yx}(\mu_0 H)$ become positive in all fields, signifying the hole-type carriers dominate the electrical transport. To better understand the compensated nature of electronic transport, two carrier model is used to fit the Hall conductivity.[37,38]

$$\sigma_{xy}(\mu_0 H) = \frac{\rho_{yx}(\mu_0 H)}{\rho_{xx}^2(\mu_0 H) + \rho_{yx}^2(\mu_0 H)} = \left[\frac{n_h \mu_h^2}{1 + (\mu_h \mu_0 H)^2} - \frac{n_e \mu_e^2}{1 + (\mu_e \mu_0 H)^2}\right] e\mu_0 H$$

where $\mu_e$ ($\mu_h$), $n_e$ ($n_h$) correspond to the mobility and density of the electron (hole). Within this model, the zero-field resistivity $\rho_{xx}(0)$ is related to the carrier concentration and mobility through the following equation $\rho_{xx}(0) = \frac{1}{e}\frac{1}{n_h \mu_h + n_e \mu_e}$.[39] Combined with this limited condition, field dependence of $\sigma_{xy}(\mu_0 H)$ is fitted, as typical example, the fitting results at 2 K are shown in the inset of Figure 3b, the well-fitting of Hall conductivity verifies the reliability of this model. The extracted $\mu_h$, $\mu_e$, $n_h$, $n_e$ as function of temperature are shown in Figure 3b,c, the carrier densities at 2 K reach $n_h = 1.5 \times 10^{19} \text{cm}^{-3}$ and $n_e = 3.63 \times 10^{19} \text{cm}^{-3}$ with carrier mobilities $\mu_h = 9735 \text{ cm}^2 \cdot \text{V}^{-1} \cdot \text{s}^{-1}$ and $\mu_e = 2195 \text{ cm}^2 \cdot \text{V}^{-1} \cdot \text{s}^{-1}$. The low carrier density in the order of $10^{19} \text{cm}^{-3}$ and high carrier mobility in all temperatures support SmAlSi as a semimetal, and the carrier mobility is comparable with the other topological semimetals, such as GdPtBi,[40] YbMnSb$_2$[41] and NbSb$_2$.[42] Another important feature is that, both $\mu_e$ and $n_e$ display remarkable change around $T_N$, signifying that the electronic states sensitive to magnetic ordering of Sm moments.

To gain insight into the electronic band structure of SmAlSi, we perform the analysis of quantum SdH oscillations. The field dependence of out-plane ($\mu_0H$ // $c$) and in-plane ($\mu_0H$ // $b$) resistivity are shown in Figure 4a,b, respectively. Strong SdH oscillations are observed and remain discernible above 20 K, and the pronounced SdH oscillations started at low field (~3 T) point out the high quality of the SmAlSi crystal. After subtracting the smooth background, the oscillatory components of $\Delta\rho_{xx}^c$ and $\Delta\rho_{xx}^b$ versus $1/\mu_0H$ at different temperatures are shown in Figure 4c,d. Furthermore, from the Fast Fourier Transform analysis (FFT) of SdH oscillations, two fundamental frequencies ($F_\alpha^c = 17.4\ T$, $F_\beta^c = 43.7\ T$ and $F_\alpha^b = 18.3\ T$, $F_\beta^b = 50.9\ T$) are clearly identified for field along $c$ and $b$ axis, indicating the presence of at least two Fermi surface pockets at the Fermi level, as shown in Figure 4e,f. According to the Onsager relation $F = (\hbar/2\pi e)A_F$, the calculated cross-sectional area of Fermi surface $A_F$ are 0.0016 Å$^{-2}$, 0.0041 Å$^{-2}$ and 0.0017 Å$^{-2}$, 0.0047 Å$^{-2}$ related to these frequencies $F_\alpha^c$, $F_\beta^c$ and $F_\alpha^b$, $F_\beta^b$. Further analysis on temperature dependence of oscillation frequencies (see Figure 4e,f and Figure S4 in supporting information), we can find the oscillation frequency of $F_\beta^c$ changes from 41 T at 10 K to 43.7 T at 2 K, indicative of the variation of Fermi surface with temperatures. Since the similar phenomena have been observed in isostructural magnetic PrAlSi,[23] not in nonmagnetic LaAlX(X=Si,Ge),[19] the change of oscillation frequency should be correlated to the magnetic ordering of Sm moments at $T_N$~11 K, and this is also corroborated by the variation of mobility and carrier density near $T_N$ as shown in Figure 3. The Fermi wave vector $k_F = (A_F/\pi)^{1/2}$, Fermi velocity $v_F = \hbar k_F/m^*$ and Fermi energy $E_F = m^*v_F^2$ could also be estimated in case of linear energy dispersion, the results are summarized in Table 1. Additionally, the amplitude of SdH oscillations can be described by the Lifshitz-Kosevich (LK) formula: [43–45]

$$\frac{\Delta\rho}{\rho_0} = \frac{5}{2}\left(\frac{\mu_0H}{2F}\right)^{1/2} \frac{2\pi^2 k_B m^*T/\mu_0 He\hbar}{\sinh(2\pi^2 k_B m^*T/\mu_0 He\hbar)}/e^{-2\pi^2 k_B T_D m^*/\mu_0 He\hbar}$$
$$\times \cos\left[2\pi\left(\frac{F}{\mu_0 H} + \frac{1}{2} - \frac{\phi_B}{2\pi} + \delta\right)\right].$$

Here, $m^*$ denotes the cyclotron mass of the carrier, $T_D$ denotes the Dingle temperature, $\phi_B$ denotes the Berry phase, $\delta$ denotes the phase factor. The value of $\delta$ depends on the dimensionality of Fermi surface and takes the value 0 or $\pm 1/8$ for the 2D and 3D systems, respectively. The thermal damping factor can be used to determine $m^*$ from the LK formula. As shown in the insets of panels e and f of Figure 4, temperature dependence of relative FFT peak amplitude can be well fitted, and the extracted $m^*$ are $m_\alpha^* = 0.1\ m_e$, $m_\beta^* = 0.07\ m_e$ for $\mu_0H$ // $c$ and $m_\alpha^* = 0.07\ m_e$, $m_\beta^* = 0.06\ m_e$ for $\mu_0H$ // $b$, where $m_e$ is the free electron mass. Similar fitting results were obtained from the high field SdH oscillations up to 48 T (see Figure S3 in the Supporting Information for details), the small cyclotron mass is comparable with the value of NbP[46] and YbMnSb$_2$.[41]

Despite light cyclotron mass and high mobility as typical characteristics of existence of Dirac or Weyl fermions, nontrivial Berry phase $\phi_B$ can be extracted from the quantum SdH oscillations and considered to be its key feature. Generally, $\phi_B$ should be zero for non-relativistic system and finite value $\pi$ for topological materials with linear dispersion. Two approaches can be used to extract the $\phi_B$. One is to fit the SdH oscillation by the LK formula directly,[43-45,47] this way is usually used to evaluate the multi-frequency oscillations when the individual peak of frequencies can't be separated. Another is to map the Landau level (LL) fan diagram where $\phi_B$ can be extracted from the intercept of linear extrapolation of LL index (*N*) to zero of inverse field 1/$\mu_0 H$,[48,49] because *N* is related to 1/$\mu_0 H$ by Lifshitz-Onsager quantization rule $(\hbar/\mu_0 H 2\pi e)A_F = N + 1/2 - \frac{\phi_B}{2\pi} + \delta$. For SmAlSi, the oscillation peaks may not be accurately determined by the LL indices using low field $\Delta\rho_{xx}$ ($\mu_0 H$ < 9 T) due to the wave superposition. In this case, the first way is used to determine $\phi_B$ based on the low field SdH oscillations. Considering that two fundamental frequencies are identified for both $\mu_0 H$ // *c* and $\mu_0 H$ // *b*, the total oscillations are fitted based on two Fermi pockets where $m^*$ are fixed to the values obtained from temperature dependent amplitude of FFT. As shown in the Figure 4g,h, two-band LK formula reproduces the resistivity oscillations well at 2 K, the yielded Berry phases are 0.62 $\pi$, 0.76 $\pi$ and 0.6 $\pi$, 0.74 $\pi$ for $F_\alpha^c$, $F_\beta^c$ and $F_\alpha^b$, $F_\beta^b$, respectively. Both Fermi pockets exhibit nontrivial Berry phases. Additionally, the Dingle temperature $T_D$ related to quantum lifetime by $\tau_q = \hbar/(2\pi k_B T_D)$ was obtained as listed in Table 1. High field SdH oscillations are desirable to detect the smaller LL index *N*, the extracted $\phi_B$ from extrapolation of high-field data is expected. Then, we analyzed the oscillatory components of $\Delta R_{xx}^c$ versus 1/$\mu_0 H$ with field up to ~48 T. As shown in Figure 5a, the peak positions of $F_\beta^c$ pocket marked by red dashed lines can be clearly resolved corresponding to the integer value of *N* shown in Figure 5b, the small index *N*=3 let the extrapolation is reliable. Under high field (1/$\mu_0 H$ < 0.04 T$^{-1}$), the SdH oscillations exhibit complex behaviors, which can be from the other undetected frequencies or Zeeman splitting effects, future study is needed to clarify its origin.

According to the LK quantization rule,[48,49] the intercept for $F_\beta^c$ pocket is determined to 0.095, within $\pm 1/8$ taking into account $\delta$, reveals the existence of nontrivial $\pi$ Berry phase in consistent with the results of LK method. In Figure S4 (in the Supporting Information), the fitting of the LL index for both 2 K and 10 K reveals the Berry phases show slight temperature dependences. It is also noted that, the de Hass van Alphen (dHvA) oscillations were detected from isothermal magnetizations at 2 K (see Figure S2 in the Supporting Information), where the analysis of dHvA oscillations give nonzero intercept 0.07 for $F_\beta^c$ pocket close to the value obtained from SdH oscillations. Thus, the nontrivial Berry phases support the presence of nontrivial topological states in SmAlSi.

To reveal the anisotropic behavior of Fermi surface, angle-dependent MRs were measured under field is rotated within the *bc*-plane and *ac*-plane, the schematic configurations are shown in the insets of Figure 6c. The obtained FFT spectra of SdH oscillations at different angles ($\theta,\varphi$) were shown in Figure 6a,b. For both rotation configurations, we can find the FFT spectra evolve systematically with similar trend as increased angles. Specifically, $F_\beta$ shows a nonmonotonic variation as increased angles reaching maximum at 60°, above that $F_\alpha$ and $F_{2\beta}$ become indistinguishable as field rotated within *ac* plane. The angular dependence of major fundamental frequencies are summarized in Figure 6c. The oscillation frequency $F_\beta$ follows $F(\theta\ or\ \varphi) = F(0)/\cos(\theta\ or\ \varphi)$ as rotating field at low angles, which unveils the FS responsible for SdH oscillation has the 2D-like features.

To better understand the electronic band topology of SmAlSi, first-principle calculations were performed to obtain the electronic band structure. The calculation is started from the nonmagnetic case with and without spin-orbit coupling (SOC). In this case, the Sm *f* electrons are kept in the core. The band structure along high symmetric lines without SOC effect is illustrated in Figure 7a, the energy dispersions around $E_F$ have several band crossings with linear dispersion characteristics around touching points. With the inclusion of SOC, the linear band crossing pionts are gaped out, and Weyl nodes may emerge in the vicinity, as shown in Figure 7b. Additionally, the calculated band structure in absence of SOC reveals that electron and hole pockets coexist at the Fermi surface (see Figure 7f), in agreement with two types of carriers revealed by Hall resistivity. In Figure 7f, the crossing between conduction and valence bands forms four closed nodal lines on the $k_x = 0$ and $k_y = 0$ mirror planes. Then, the electronic structures of SmAlSi in magnetic-ordered state are calculated, in which Sm $f$ electrons are put in the valence and a Hubbard energy $U$ of 6.4 eV was used in the calculation. Considering the FM state, the calculations reveal SmAlSi has magnetic moment of 1.03 $\mu_B$/Sm in close to the experimental value ~0.75 $\mu_B$/Sm. As displayed in Figure 7c, the Sm $f$ orbital is partially occupied, giving rise to the local magnetism of SmAlSi. In magnetic ordered state, the band structures without SOC and with SOC are illustrated in panels d and e of Figure 7. Compared to the nonmagnetic case, the spin-up and spin-down sub-bands without SOC split in the FM state, while the band structure is slightly changed, indicating that the magnetism can shift the location of Weyl nodes in momentum space. By including SOC, the energy dispersions and positions near the crossing points become more complex and tuned, as example, the Weyl cone near N-$\Sigma_1$ and $\Sigma_1$-Zpoints are tilted indicative of possible type II Weyl states (see Figure 7e). These calculated results share some similarities with its isostructural RAlX (R = La, Ce, Pr) compounds identified as WSM materials,[20,24-26] the Weyl nodes stem from the broken inversion symmetry and magnetism in this family. In combination with the light cyclotron mass and nontrivial Berry phase from experimental results, SmAlSi can be served as a new magnetic WSM system for exploiting the interplay between magnetism and Weyl states.

In summary, we have grown high quality single crystals of SmAlSi and performed systematical magnetotransport studies. High field magnetotransport reveals that SmAlSi exhibit large non-saturated MRs ~5200% at 2 K under 48 T, accompanied by a linear field dependent MR behavior for $\mu_0H$ > 18 T. The analysis of SdH oscillations reveals the existence of two FS pockets with nonzero π Berry phase, light cyclotron mass and high carrier mobility as typical features of Weyl fermions in SmAlSi, in agreement with the electronic band structure calculations. The results reveal the high quality SmAlSi as an interesting material on understanding WSM physics in RAX family.

## ■ ACKNOWLEDGMENTS

We acknowledge financial support from the National Natural Science Foundation of China (grant no. 11874158 and grant no. 12004123) and the Fundamental Research Funds for the Central Universities (grant no. 2019KFYXKJC008). We would like to thank the staff of the analysis centre of Huazhong University of Science and Technology for their assistance in structural characterizations.

# Figure captions

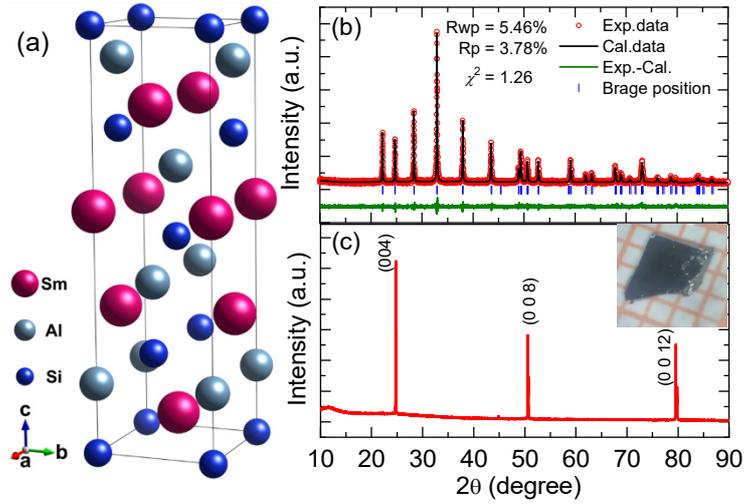

Figure 1. (a) Crystal structure of SmAlSi. (b) The experimental and refined powder XRD spectra. (c) Singe crystal XRD patterns of (001) plane, inset shows the optical image of single crystal.

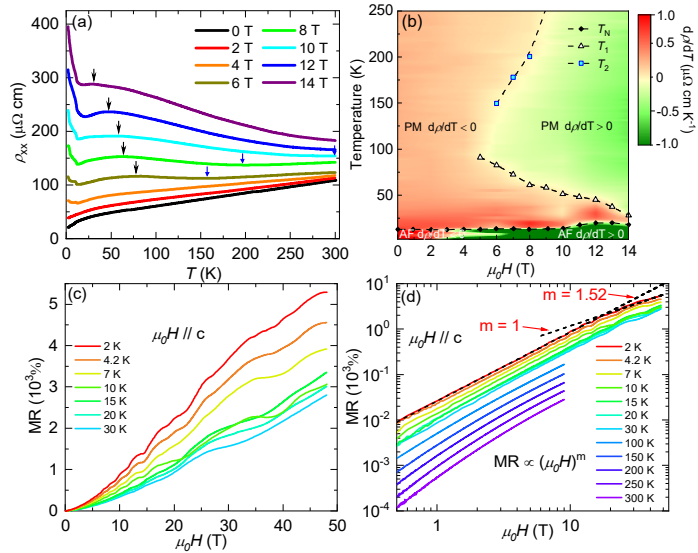

Figure 2. (a) Temperature dependence of resistivity $\rho(T)$ for $\mu_0H // c$. The humps ($T_1$) and valleys ($T_2$) are marked by the black and blue arrows, respectively. (b) Temperature-field phase diagram of SmAlSi. (c) Magnetic field dependent MRs at different temperatures for $\mu_0H // c$. (d) A double-logarithmic plot of MRs. The two black dashed fitting lines show the different slopes of MR at low and high fields (m = 1.52 in low fields and m = 1 in high fields).

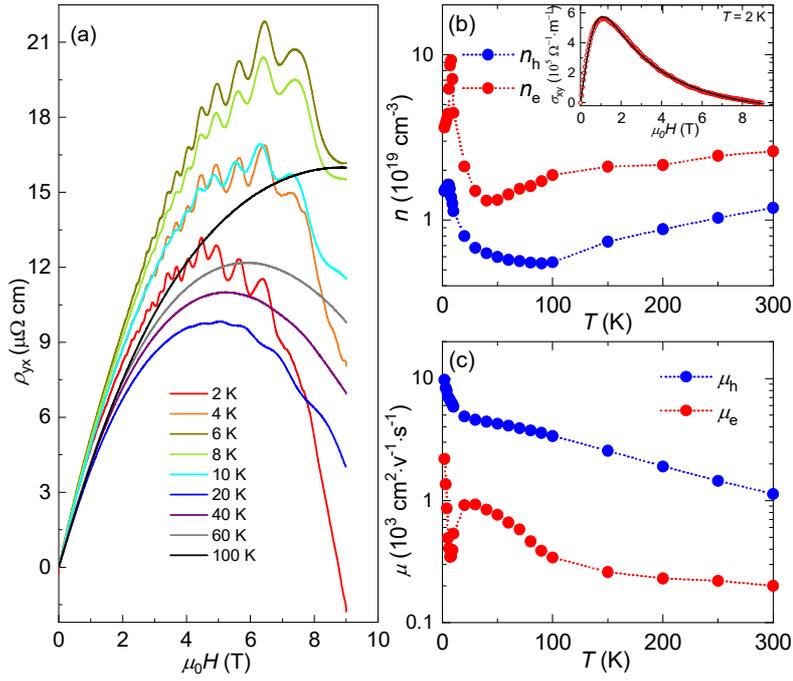

Figure 3. (a) Field dependence of Hall resistivity $\rho_{yx}(\mu_0 H)$ of SmAlSi at different temperatures. (b and c) Temperature dependence of carrier density and carrier mobility, respectively. The inset of (b) displays the experimental and fitted results of $\sigma_{xy}(\mu_0 H)$ at 2 K.

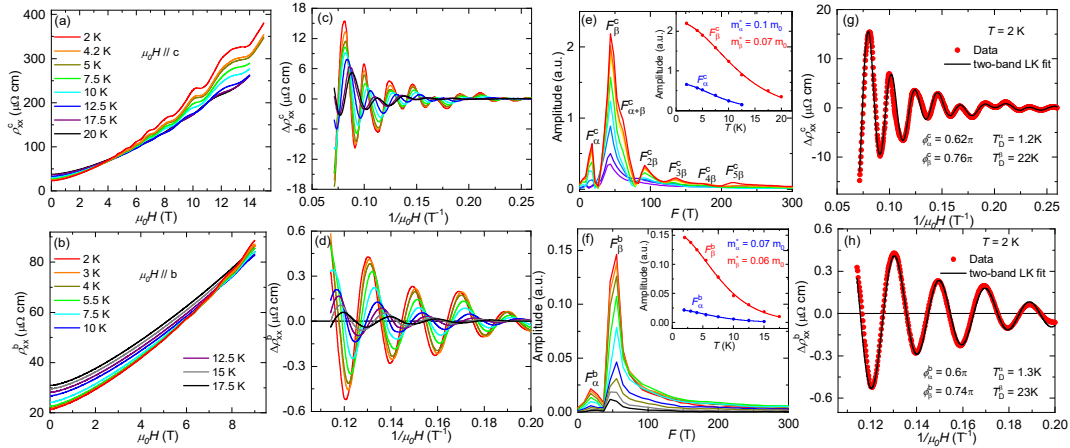

Figure 4. The analysis of SdH oscillations of SmAlSi for $\mu_0 H$ // c axis (a,c,e,g) and $\mu_0 H$ // b axis (b,d,f,h). (a and b) Field dependence of resistivity $\rho_{xx}(\mu_0 H)$ at different temperatures. (c and d) $\Delta\rho_{xx}$ versus $1/\mu_0 H$ at different temperatures. (e and f) The FFT spectra of oscillations with two fundamental frequencies. Insets: Temperature dependence of FFT peak amplitude fit by the Lifshiz-Kosevich (LK) formula. (g and h) The LK fit (black line) of the oscillation pattern (red points) at 2 K.

Table 1. Parameters derived from SdH oscillations for SmAlSi. $F$, oscillation frequency; $A_F$, external cross-sectional area of the FS; $K_F$, Fermi vector; $v_F$, Fermi velocity; $E_F$, Fermi energy; $T_D$, Dingle temperature; $\tau_q$, quantum relaxation time; $m^*/m_e$, effective mass; $\phi_B$, Berry phase.

| | $F$(T) | $A_F$(Å$^{-2}$) | Fraction | $K_F$(Å$^{-1}$) | $v_F$($10^5$m/s) | $E_F$(mev) | $T_D$(K) | $\tau_q$($10^{-13}$ s) | m*/m$_e$ | $\phi_B$(LK fit, $\delta = \pm 1/8$) |
|---|---|---|---|---|---|---|---|---|---|---|
| B//c | 17.4 | 0.0016 | 0.072% | 0.022 | 2.6 | 37.5 | 1.2 | 9.9 | 0.10 | [0.62±0.25]π |
| | 43.7 | 0.0041 | 0.18% | 0.036 | 6.1 | 140 | 22 | 0.54 | 0.07 | [0.76±0.25]π |
| B//b | 18.3 | 0.0017 | 0.22% | 0.023 | 3.9 | 58.8 | 1.3 | 9.2 | 0.07 | [0.60±0.25]π |
| | 50.9 | 0.0047 | 0.70% | 0.039 | 7.8 | 202 | 23 | 0.52 | 0.06 | [0.74±0.25]π |

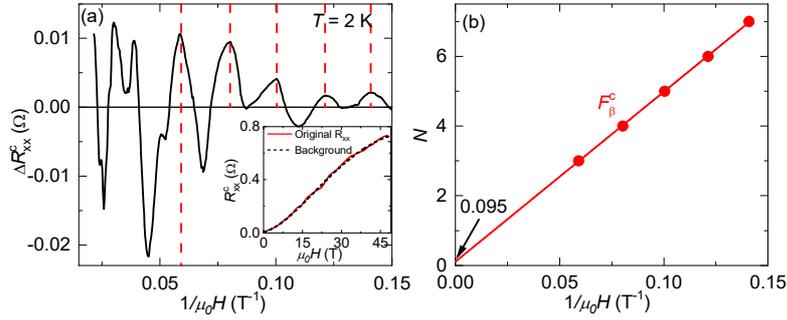

Figure 5. (a) The oscillatory components $\Delta R_{xx}$ at 2 K for $\mu_0 H$ // c. Red dashed lines represent the positions of peaks. The inset shows magnetic-field dependence of original $R_{xx}$ (red line) and the background of a polynomial fit (dashed black line) at 2 K for $\mu_0 H$ // c. (b) The LL fan diagram for the $F_\beta^c$ pocket at 2 K.

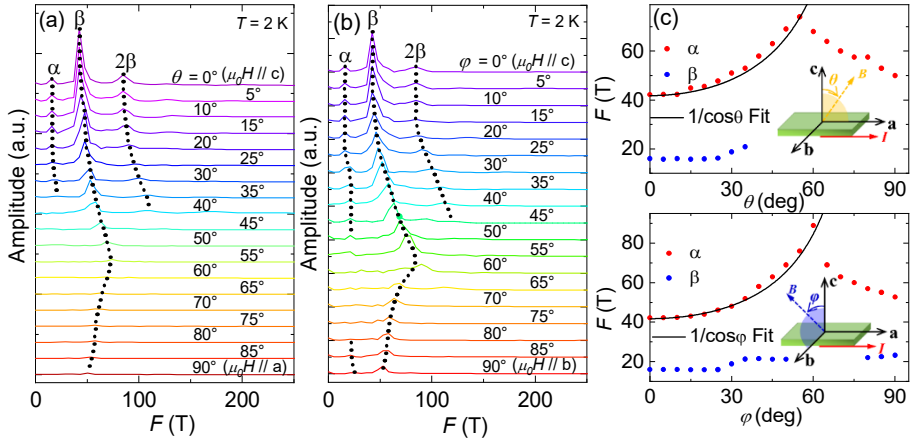

Figure 6. (a,b) The angular dependence of FFT spectra obtained from SdH oscillations for field rotating within the ac-plane and bc-plane, respectively. The dashed lines are guides to the eyes. (c) The angular dependence of frequency of oscillations for the α and β pockets. Insets: configurations for measurement on angle dependent resistivity at 2 K, where θ or φ is defined as the angle between field and c axis.

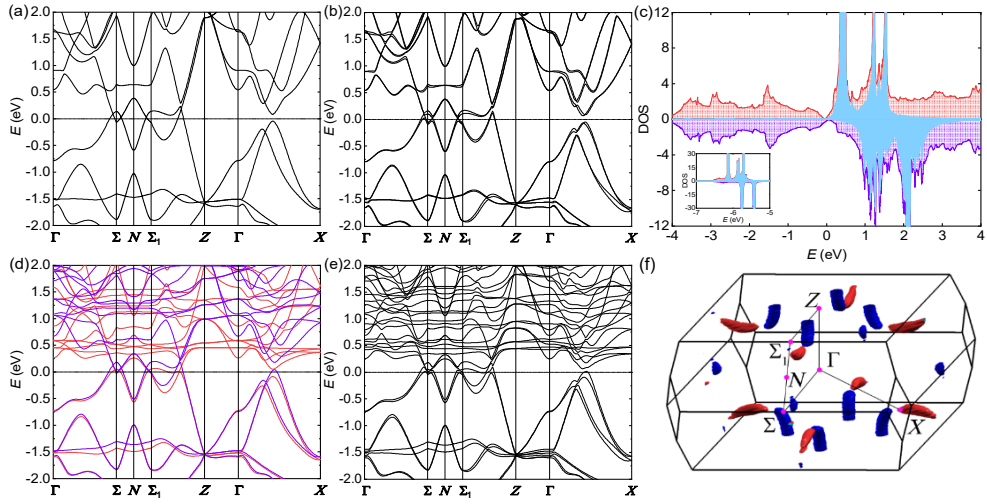

Figure 7. Band structure for SmAlSi in nonmagnetic state (a) without SOC and (b) with SOC. (c) Calculated DOS of SmAlSi in FM state. The spin-up and spin-down partial DOS are plotted in red and purple colors, respectively. The 4f states of Sm are represented by blue shaded area. The inset shows the Sm f states below EF. (d) The band structure of SmAlSi along high symmetry lines without SOC in FM state. The red and purple lines represent the spin-up and spin-down sub-bands in the FM state, respectively. (e) The band structure of SmAlSi in the FM state with SOC. (f) Electron (blue) and hole (red) pockets are shown in the Brillouin zone (BZ) without SOC in nonmagnetic state.